\pdfoutput=1
\documentclass[twocolumn, pra, floatfix, superscriptaddress, showpacs]{revtex4-1}
\usepackage{amsmath,mathrsfs,bbm,graphicx,natbib}
\usepackage[usenames,dvipsnames]{color}
\usepackage[colorlinks=true,
            urlcolor=black,
            citecolor=black,
            linkcolor=black]{hyperref}
\usepackage{Definitions}

\begin{document}
\title{Single Shot Quantum State Estimation via a Continuous Measurement in the Strong Backaction Regime}
\author{Robert L. Cook}
\affiliation{Center for Quantum Information and Control, University of New Mexico, Albuquerque, NM 87131, USA}
\author{{ Carlos A. Riofr\'{i}o}}
\affiliation{Dahlem Center for Complex Quantum Systems, Freie Universit\"at Berlin, 14195 Berlin, Germany}
\affiliation{Center for Quantum Information and Control, University of New Mexico, Albuquerque, NM 87131, USA}
\author{{ Ivan  H. Deutsch}}
\affiliation{Center for Quantum Information and Control, University of New Mexico, Albuquerque, NM 87131, USA}
\date{\today}

\begin{abstract}
    We study quantum tomography based on a stochastic continuous-time measurement record obtained from a probe field collectively interacting with an ensemble of identically prepared systems.  In comparison to previous studies, we consider here the case in which the measurement-induced backaction has a nonnegligible effect on the dynamical evolution of the ensemble.  We formulate a maximum likelihood estimate for the initial quantum state given only a single instance of the continuous diffusive measurement record.  We apply our estimator to the simplest problem -- state tomography of a single pure qubit, which, during the course of the measurement, is also subjected to dynamical control.  We identify a regime where the many-body system is well approximated at all times by a separable pure spin coherent state, whose Bloch vector undergoes a conditional stochastic evolution.   We simulate the results of our estimator and show that we can achieve close to the upper bound of fidelity set by the optimal POVM.  This estimate is compared to, and significantly outperforms, an equivalent estimator that ignores measurement backaction.
\end{abstract}


\maketitle

\section{Introduction}
A fundamental task in quantum information processing is the ability to both reliably prepare an arbitrary quantum state and experimentally verify its production.  Traditional quantum state tomography (QST) relies on an exhaustive procedure where the target state is repeatedly prepared and then destructively measured in an informationally complete number of measurement settings.  Such a procedure is often extremely time intensive, requiring both a significant amount of data and  post-processing time \citep{haffner_scalable_2005, leibfried_creation_2005}.

These inefficiencies can be significantly reduced when one can perform a weak continuous measurement, acting collectively on an identically prepared ensemble, in conjunction with well chosen dynamical control~\cite{silberfarb_quantum_2005, riofrio_quantum_2011}.  In particular, consider an ensemble of $N$ systems prepared in an identical tensor product state $\rho_{tot} = \rho_0^{\otimes N}$, experiencing a known time-dependent control Hamiltonian while simultaneously coupled to a traveling wave probe.  If the control drives the system such that a continuous measurement of the probe is informationally complete, then one can use this measurement record to obtain a high-fidelity estimate of the initial state of the system, $\rho_0$.

This protocol has been implemented in experiments~\cite{smith_efficient_2006, smith_quantum_2013} with quantum states encoded in the hyperfine spins of an ensemble of laser-cooled cesium atoms controlled with magneto-optical fields~\cite{chaudhury_quantum_2007, merkel_quantum_2008, smith_quantum_2013} and measured with polarization spectroscopy~\cite{smith_continuous_2004}.  By applying an appropriate estimator to the measurement record, one can obtain high-fidelity reconstructions of arbitrary states in the 16-dimensional hyperfine ground state manifold of cesium. However, these experiments were performed far from idealized conditions.  The reconstructions were ultimately limited by systematic errors and decoherence caused by spontaneous emission.  While detrimental to the final fidelity, these limitations simplified the analysis, as the collective effects of quantum backaction were completely negligible.  Because of these facts, any fundamental limits of continuous measurement based QST have yet to be addressed.

Here, we extend this protocol to an idealized regime, free from technical imperfections and decoherence, where any limitations are solely due to the quantum backaction induced by the measurement itself and thus fundamental to the tomographic protocol.  The primary effects of measurement backaction are to introduce correlations between the atoms, \emph{i.e.} spin squeezing~\cite{hammerer_quantum_2010}, as well as to perturb the mean spin in a random and nonlinear way.  Both effects greatly increase the complexity of the problem, as the former necessitates a many-body description and the latter prevents the use of many standard tomographic techniques, \emph{e.g.}, convex optimization.  This work addresses these issues by deriving a general likelihood function for a continuous-time diffusive measurement of a collective spin projection, and derives an efficiently computable approximation in the case of pure qubits.  We then use this function to numerically compute a maximum likelihood estimate (MLE) to reconstruct the initial state.  We compare our results to the well-known bounds for the average fidelity~\cite{massar_optimal_1995}, which is achieved by the optimal collective POVM ~\cite{bagan_comprehensive_2005}.

The remainder of this paper is structured as follows.  We first establish a general mathematical model for a continuous-time, collective-spin measurement via polarization spectroscopy, with particular emphasis on the conditions under which quantum backaction cannot be neglected.  We then derive a maximum likelihood estimator for the initial state of the ensemble given a diffusive continuous-time measurement.  From the general expression, we specialize to estimating the initial state of a pure qubit given an ensemble of identical copies.  We derive an efficiently computable approximation to the exact expression under the condition that the entangling effects of the measurement backaction are negligible, while stochastic kicks to the Bloch vector induced by the measurement remain important.   We then numerically test the performance of the approximate maximum likelihood estimate for a moderate number of qubits and compare the results both to the optimal POVM for quantum tomography and an estimator that completely ignores the effect of measurement backaction.  We conclude with a summary and outlook for future studies.

\section{Spin estimation through polarization spectrocopy}
\begin{figure}[t]
    \begin{center}
        \includegraphics[width=\hsize]{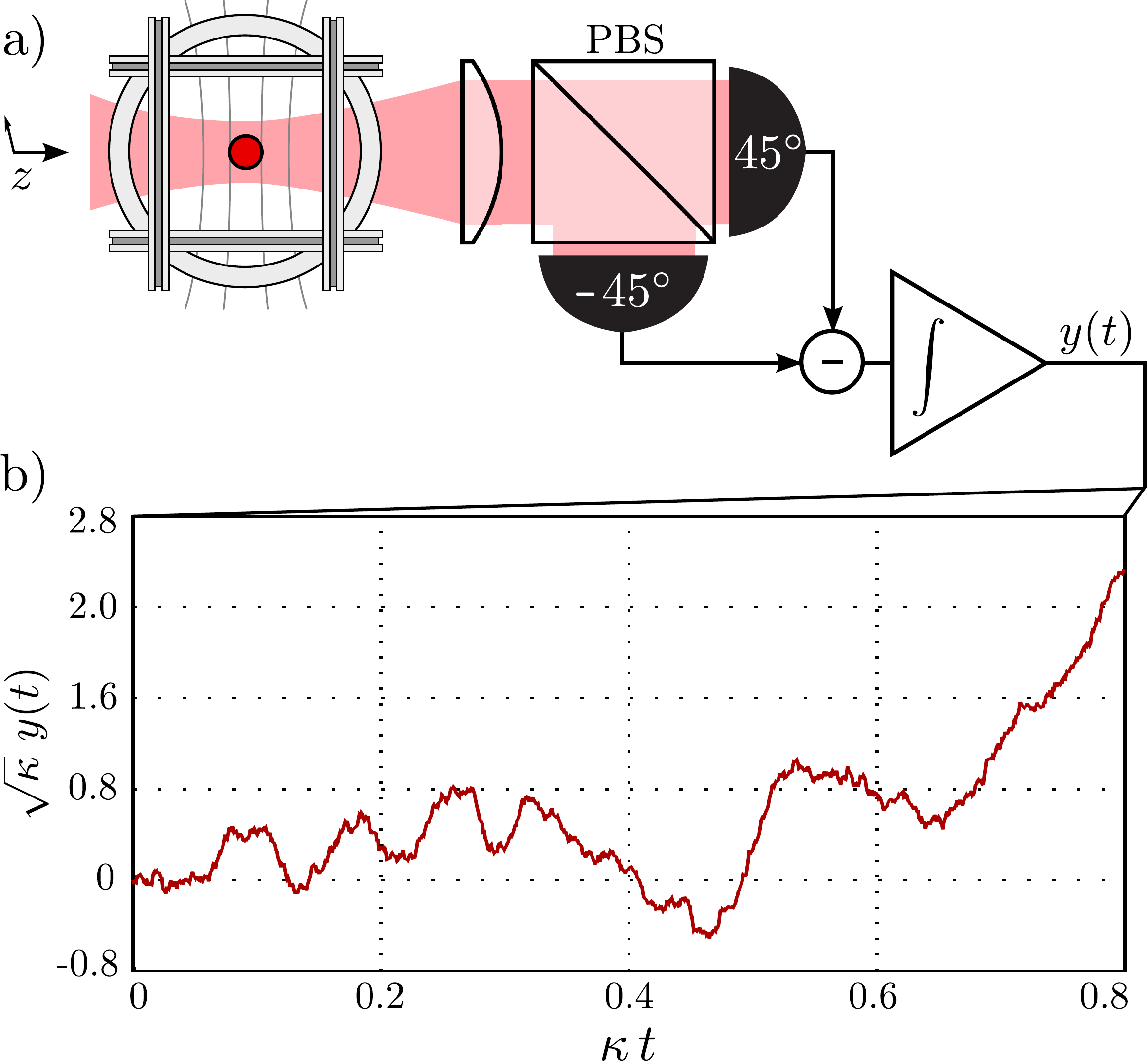}
        \caption{\label{fig:Schematic} (Color online) Schematic and Sample Measurement. a) An atomic ensemble is probed by an off-resonant, linearly polarized laser, while simultaneously subjected to external RF magnetic control fields.  The outgoing laser is measured by a balanced polarimeter, whose integrated current generates a noisy measurement record $y(t)$.  b) A typical simulated measurement record for $N = 50$ symmetrically coupled qubits initialized in a spin coherent state along the $x$ axis.}
	\end{center}
\end{figure}
We consider cold atomic spins measured via polarization spectroscopy as our model platform in which to examine the fundamental limits of QST based on continuous measurement and control~\cite{deutsch_quantum_2010}; a schematic is shown in Fig. \ref{fig:Schematic}.  The measurement is made via the Faraday interaction, whereby the linear polarization of an off-resonant probe laser rotates in proportion to the collective magnetization of the atomic ensemble along the direction of propagation of the probe.  For a system composed of $N$ atoms identically coupled to the probe field, a measurement of this rotation results in a quantum non-demolition (QND) measurement of the collective angular momentum operator, $J_z = \sum_{i=1}^N j_z^{(i)}$, where $j_z^{(i)}$ is the $z$--axis projection of the $i^{th}$ atomic spin operator.  This measurement occurs at a rate $\kappa$, which is set by the input photon flux times the rate at which a single atom will scatter an incident photon into the orthogonal polarization mode.

A balanced polarimeter measuring in a basis 45$^\circ$ to the input polarization implements an effective homodyne measurement, where the probe field acts as the local oscillator~\cite{baragiola_three-dimensional_2014}.  A continuous record of the integrated photocurrent is described by a stochastic process $\set{y(t) : 0 \le t \le T }$, where $T$ is the fixed final time.  For a system prepared in the definite initial condition $\rho_{tot}(0)$, this can be be written as~\cite{bouten_introduction_2007}
\begin{equation}\label{eq:integratedMeasurement}
y(t) = w(t) + \sqrt{\kappa} \int_0^t \Tr \left( \rho_{tot}(s) J_z\right)\,ds ,
\end{equation}
where $\set{w(t) : t \ge 0 }$ is a realization of the Wiener process, representing the time integral over the uncorrelated shot noise introduced by the quantum limited measurements made at every time $t$ (see Fig. \ref{fig:Schematic}b).  In a single run of the experiment, the evolution of $\rho_{tot}(t)$ conditioned on this measurement record is governed by the well known diffusive stochastic master equation (SME)~\cite{jacobs_straightforward_2006, wiseman_quantum_2010} ($\hbar = 1$),
\begin{equation}\label{eq:SME}
  \begin{split}
    d\rho_{tot}(t) &= -i[H_c(t),\rho_{tot}(t)]\,dt + \gamma_{diss}\, \mathcal{D}[\rho_{tot}(t)]\, dt \\
     &\quad + \frac{\kappa}{4} \mathcal{L}[ \rho_{tot}(t) ]\, dt  +  \frac{\sqrt{\kappa}}{2} \mathcal{H}[ \rho_{tot}(t) ]\, dv(t), \\
    dv(t)&\equiv dy(t) - \sqrt{\kappa} \Tr\big(\rho_{tot}(t) J_z\big)\, dt,
  \end{split}
\end{equation}
where $H_c(t)$ is the externally applied control Hamiltonian and we have defined the maps
\begin{equation}
  \begin{split}
  \mathcal{L}[ \rho_{tot} ] &\equiv J_z\,\rho_{tot}\,J_z  - \half J_z^2\,\rho_{tot} - \half \rho_{tot} J_z^2, \\
  \mathcal{H}[ \rho_{tot} ] &\equiv J_z\,\rho_{tot} + \rho_{tot}\,J_z -  2 \Tr\big( \rho_{tot}(t) J_z\big)\, \rho_{tot}.
  \end{split}
\end{equation}
We have also included an additional general channel $\mathcal{D}[\rho_{tot}]$ that accounts for any additional sources of decoherence occurring at a characteristic rate $\gamma_{diss}$.  The stochastic process defined by the differential $dv(t)$ is known as the innovation process and determines the strength of the measurement backaction in a given interval.   When the measurement record is consistent with the true state of the system, the innovation is a Wiener process, $dv(t) = dw(t)$, as follows from inverting Eq. (\ref{eq:integratedMeasurement}).  However, the task of QST is to estimate an \emph{unknown} quantum state given some data.  To derive such an estimate, one must evolve a conditional state from an initial condition $\rho'(0)$ not equal to the initial condition used to generate the data.  Written in terms of the innovation, Eq. (\ref{eq:SME}) is still valid, but in general, we cannot assume that the innovation is a Wiener process.

Previous experiments on QST via continuous measurement~\cite{smith_efficient_2006, smith_quantum_2013} operated in a regime where the control Hamiltonian $H_c(t)$ and decoherence rate $\gamma_{diss}$ were much larger in magnitude than the measurement terms proportional to $\kappa$.  The measurement duration, $T$, was chosen such that $\gamma_{diss}T<1$.  Thus, since $\kappa T \ll 1$, the stochastic measurement outcomes in the measurement record  are completely dominated by the shot noise in the probe rather than  the ``projection noise'' uncertainty of the state, $(\Delta J_z)^2_{PN} =  {\rm Tr}\left( \rho_{tot}(0) J_z^2\right)- {\rm Tr}\left( \rho_{tot}(0) J_z\right)^2$.  In that case, measurement backaction is negligible over the duration of the measurement and the system will remain unentangled.  The expected value of the collective spin is then well approximated as $\Tr\left( J_z \rho_{tot}(t) \right) \approx N\Tr\left( j_z \tilde{\rho}(t) \right)$, where $\tilde{\rho}$ is the single particle density operator that evolves solely under an unconditional master equation
\begin{equation}\label{eq:singleAtomMasterEq}
  \tfrac{d}{dt} \tilde{\rho}(t) = -i [h_c(t), \tilde{\rho}(t)] + \gamma_{diss}\mathcal{D}^{(1)}(\tilde{\rho}(t)),
\end{equation}
where $h_c(t)$ is the single atom control Hamiltonian, and $\mathcal{D}^{(1)}$ is the single atom decoherence map.  In this backaction-free approximation, the tomographic estimate for the initial state $\tilde{\rho}(0)$ reduces to a standard problem of constrained maximum likelihood~\cite{riofrio_quantum_2011}.

Here we consider the opposite regime, where $\gamma_{diss} = 0$ and $\kappa T$ is not necessarily small.  This  presents a formidable challenge due to the nonlinear nature of $\mathcal{H}[\rho_{tot}]$, as well as the fact that the future values of $y(t)$ depend on its past through the conditional nature of  $\rho_{tot}$.  For simplicity, we restrict our attention here to the case of pure-state, nondisspative dynamics.  When $\gamma_{diss} = 0$ and assuming perfect measurement (i.e., unit quantum efficiency), the evolution of an initial pure state will remain pure.  It is then sufficient to propagate a collective state vector, $\ket{\Psi(t)}$, which evolves according to a conditional Schr\"{o}dinger equation (CSE)~\cite{jacobs_straightforward_2006}
\begin{equation}\label{eq:CSE}
\begin{split}
  d \ket{\Psi(t)} &= \Big[ -i H_c(t) - \tfrac{1}{8} \kappa \big(J_z  - \expect{J_z}_{\Psi(t)} \big)^2\, \Big] \ket{\Psi(t)}\, dt\\
   &\quad + \half \sqrt{\kappa}\,\big( J_z - \expect{J_z}_{\Psi(t)} \big)\, \ket{\Psi(t)}\,  d v(t), \\
    dv(t) &= dy(t) - \sqrt{\kappa} \expect{J_z}_{\Psi(t)}\, dt,
\end{split}
\end{equation}
where $\expect{J_z}_{\Psi(t)} = \braOket{\Psi(t)}{J_z}{\Psi(t)}$.  Our goal is to deduce the initial state $\ket{\psi(0)}$ of one member of an identical ensemble, given an initial product state, $\ket{\Psi(0)}=\ket{\psi(0)}^{\otimes N}$, and a continuous measurement record of the form Eq. (\ref{eq:integratedMeasurement}), when the collective state evolves according to Eq. (\ref{eq:CSE}).

\section{The likelihood function}
Estimating an initial quantum state from an observed measurement record is fundamentally a problem of statistical inference.  Here we utilize a MLE given the measurement record $\set{y(t) : 0 \le t \le T}$, obtained over time from a collective measurement on a single ensemble.  Our derivation hinges on the known form of the measurement record given in Eq. (\ref{eq:integratedMeasurement}), and that the quantum trajectory is consistent with our model of homodyne detection.  Given this, we are able to apply well developed classical methods for analyzing continuous stochastic processes~\cite{liptser_statistics_2001}.

Defining a likelihood function for continuous diffusive stochastic processes is more mathematically involved than for discrete or single-valued random variables.  We begin by considering a general stochastic process, $\set{ x(t): 0 \le t \le T}$, defined by the integral
\begin{equation}\label{eq:generalDiffusion}
x(t) = w(t) + \int_0^t m\boldsymbol(\theta_0, s, x(s)\boldsymbol)\, ds,
\end{equation}
where $m\boldsymbol(\theta_0, t, x(t)\boldsymbol)$ is the integrated mean signal that is assumed to be a time-dependent functional of the past history of $\set{x(t)}$, and $\theta_0$ is a vector of unknown parameters in the model.  By assuming that $m\boldsymbol(\theta_0, t, x(t)\boldsymbol)$ can only depend on $\set{ x(s): 0 \le s < t }$, we can compute the probability density for $\set{x(t): 0 \le t \le T}$ by making a simple change of variables in the probability density for the Wiener process.  The defining properties of the Wiener process are: ($i$) it has a continuous trajectory starting from zero and $(ii)$ its increments are independent, mean zero, Gaussian distributed random variables, whose variance is equal to the increment's time duration.  These criteria imply that the density for the Wiener process is defined by a product of nested Gaussians, and therefore the density for $\set{x(t)}$ will also be given by a product of Gaussians.

This is most easily seen by first considering a countably dense set of $n$ times $\big\{ t_i \in [0, \infty):\quad 0 = t_0 < t_1 < \dots < t_n = T \big\}$ and then examining the continuous limit.  To ease the notation, we define the quantities $m_i^{\theta_0} \equiv m\boldsymbol(\theta_0, t_i, x(t_{i})\boldsymbol)$, $\Delta t_i \equiv t_{i} - t_{i-1}$, and $\Delta x_i \equiv x(t_i) - x(t_{i-1})$.  We obtain the continuous limit as $\Delta t_i \rightarrow 0$.  For simplicity, we will also assume that $n$ is large enough such that the approximation $\int_{t_{i-1}}^{t_i}  m\boldsymbol(\theta_0, s, x(s)\boldsymbol)\, ds \approx m_{i-1}^{\theta_0}\, \Delta t_i$ is valid.  The joint probability that each $x_i$ will be found in the corresponding interval $[a_i, b_i]$ is then well approximated by the integrals
\begin{multline}\label{eq:finiteDimDist}
  \mathbbm{P}\boldsymbol( \set{x_i \in [a_i, b_i]} \boldsymbol) \approx \int_{a_1}^{b_1} dx_1 \dots \int_{a_n}^{b_n} dx_n \\
 \prod_{i=1}^n\frac{\exp\left[ - \frac{1}{2 \Delta t_i} \left(\Delta x_i  - m_{i-1}^{\theta_0}\, \Delta t_i\right)^2 \right]}{ \sqrt{2 \pi \Delta t_i}}.
\end{multline}

A natural way to perform MLE would be to consider the integrand in Eq. (\ref{eq:finiteDimDist}) as the likelihood $\mathcal{L}(\theta)$, \emph{i.e.}, a function of the input parameter vector $\theta$, given an observation of $\set{x(t): 0 \le t \le T}$ as determined by the unknown parameters $\theta_0$,
\begin{equation}\label{eq:finiteLikelihood}
\mathcal{L}(\theta) =  \prod_{i=1}^n \frac{\exp\left[ - \frac{1}{2 \Delta t_i} \left(\Delta x_i - m_{i-1}^{\theta}\, \Delta t_i\right)^2 \right]}{ \sqrt{2 \pi \Delta t_i}}.
\end{equation}
However, such a likelihood fails to be of use in the continuous-time limit because as $\Delta t_i \rightarrow 0$, the measurement record is dominated by shot noise and is ultimately independent of $\theta$.  This can be seen by substituting the definition of $\Delta x_i$ from Eq. (\ref{eq:generalDiffusion}), resulting in,
\begin{equation}
\mathcal{L}(\theta) = \prod_{i = 1}^n \frac{\exp\left[ - \half  \left( \tfrac{\Delta w_i}{\sqrt{\Delta t_i}} - \big( m_{i-1}^{\theta} - m_{i-1}^{\theta_0}  \big)\sqrt{ \Delta t_i} \right)^2 \right]}{  \sqrt{2 \pi \Delta t_i}}.
\end{equation}
For any noise realization and $\Delta t_i > 0$, the random variables $\xi_i \equiv \Delta w_i/\sqrt{\Delta t_i}$ are mean zero Gaussian random variables with \emph{unit} variance.   Therefore, at every time index, an estimator maximizing Eq. (\ref{eq:finiteLikelihood}) would minimize the squared deviation of a number $\xi_i \sim O(1)$, with a mean proportional to $\sqrt{\Delta t_i}$.  In the limit $\Delta t_i \rightarrow 0$, this expression is independent of $\theta$ and depends solely on the unwanted shot noise. While we could reduce the effect of shot noise by coarse graining the measurement record over longer time intervals, such a procedure would also necessarily coarse grain over the time dependence in $m$, possibly resulting in a loss of information about $\theta_0$.

Fortunately, we can make full use of the continuous measurement record by instead considering a \emph{likelihood-ratio} between a candidate parameter $\theta_1$ and a reference parameter $\theta_2$.  By doing so the divergences represented by $\xi_i$ cancel, leaving a useful expression in the continuous-time limit.  Computing this ratio and simplifying gives
\begin{multline}
 \frac{\mathcal{L}(\theta_1)}{\mathcal{L}(\theta_2)} =  \exp\left\{ \sum_{i= 1}^n \left[ m^{\theta_1}_{i-1} - m^{\theta_2}_{i-1}\right]\, \Delta x_{i} \right. \\
 \left. - \half \sum_{i= 1}^n \left[ (m^{\theta_1}_{i-1})^2 - (m^{\theta_2}_{i-1})^2\right]\, \Delta t_{i} \right\}.
\end{multline}

The limit $\Delta t_i \rightarrow 0$ of this expression exists and is meaningful, resulting in the exponentiated It\=o integral,
\begin{multline}\label{eq:likelihoodRatio}
\hspace{-0.35 cm}\Lambda(\theta_1, \theta_2) =  \exp\left\{ \int_0^T \hspace{-0.25cm} \big[ m \boldsymbol( \theta_1, t, x(t) \boldsymbol) - m\boldsymbol(\theta_2, t, x(t)\boldsymbol)\big] d x(t) \right. \\
 \qquad \left. - \half \int_0^T \left[ m \boldsymbol( \theta_1, t, x(t) \boldsymbol)^2 - m\boldsymbol( \theta_2, t, x(t) \boldsymbol)^2 \right] dt \right\}.
\end{multline}

To turn this general expression into the form we will use, we first note that given a measurement record, $\set{y(t) : 0 \le t \le T}$, and a valid initial condition, the expectation value $\expect{J_z}_{\Psi(t)}$ can be viewed as a time-dependent functional of the measurement record.   We also note that a maximization of $\Lambda$ with respect to its first argument is equivalent to maximizing a \emph{log likelihood ratio} (LLR), $\lambda \equiv \ln\, \Lambda$.  Under the replacements $x(t) \rightarrow y(t)$, $\theta_n \rightarrow \Psi_n(0)$, and $m\boldsymbol(\theta_n, t, x(t)\boldsymbol) \rightarrow \sqrt{\kappa} \,\expect{ J_z}_{\Psi_n(t)}$, we have
\begin{equation}\label{eq:logLike}
\begin{split}
  \lambda\boldsymbol(\Psi_1(0),\Psi_2(0)\boldsymbol) &\define \ln(\Lambda)\\
   &=  \sqrt{\kappa} \int^{T}_{0} d y(s) \left( \expect{J_z}_{\Psi_1(s)} - \expect{J_z}_{\Psi_2(s)} \right) \\
   &\quad  - \frac{\kappa}{2}\int_0^T ds\, \left( \expect{J_z}_{\Psi_1(s)}^2 - \expect{J_z}_{\Psi_2(s)}^2\, \right).
\end{split}
\end{equation}
The MLE we will use is then
\begin{equation}\label{eq:maxLogLike}
  \ket{\Psi_{ML}} = \argmax_{\Psi_1}\left[  \lambda\boldsymbol(\Psi_1(0),\Psi_2(0)\boldsymbol)\, \right].
\end{equation}
In principle the exact choice of $\Psi_2(0)$ is irrelevant for computing $\ket{\Psi_{ML}}$, as the replacement $\Psi_2(0) \rightarrow \Psi_3(0)$ changes $\lambda$ by a finite additive constant, but this does not affect where the maximum occurs.  In practice, however, an initial condition that is radically different from the true one greatly reduces the numerical stability of Eq. (\ref{eq:CSE}).  This fact impacts the choice of reference and the reconstruction algorithm we implement.

\section{Estimating the state of a pure qubit}
As a first step towards understanding the fundamental limits of QST based on continuous-time measurement and control, we consider the simplest problem -- reconstructing the state of a pure single qubit. We assume that we are initially given $N$ qubits, each initialized in an unknown yet pure state $\ket{\psi_0}$.  We then assume the total evolution preserves the exchange symmetry of the system, thus allowing us to restrict our attention to states that are in the fully symmetric subspace of the many-body system.   The evolution thus preserves the total collective angular momentum quantum number at its maximum value $J = N/2$.  Therefore, instead of considering the entire $2^N$ dimensional tensor-product Hilbert space, we are able to restrict our attention to the evolution to the $d=N+1$-dimensional exchange-symmetric subspace.

A key ingredient of the protocol is to drive the system with a control Hamiltonian that ensures that the measurement record is informationally complete.  Following the work of  Riofr\'io {\em et al.}~\cite{riofrio_quantum_2011}, we choose a control Hamiltonian, $H_c(t)$, that is randomized between a set of operators that rapidly generates the group of $SU(2)$ rotations,
\begin{equation}\label{eq:Hcontrol}
H_c(t) = \V{b}(t) \cdot \V{J}=\sum_i  \V{b}(t) \cdot \V{\sigma}^{(i)}/2,
\end{equation}
with
\begin{equation}
\V{b}(t) = \frac{\pi}{2\, \tau} \sum_{i =1} \chi_{[{i-1}, i)}(\,t/\tau)\, \mathbf{e}_i.
\end{equation}
Here $\set{\mathbf{e}_i}$ are uniformly sampled directions on the unit sphere, $\tau$ is the transition period, and the indicator function $\chi_{[a,b)}(x) = 1$ for $x \in [a,b)$ and zero otherwise. The choice of a Larmor frequency $\Omega_b = \pi/(2 \tau)$ is an attempt to maximize the information gain, \emph{e.g.}, if $\mathbf{e}_i = \mathbf{e}_x$ then a $\pi/2$ rotation is needed to rotate the unobserved $\mathbf{e}_y$ component of the collective spin onto the measurement axis, $\mathbf{e}_z$.

For an arbitrary control law, the estimate $\ket{\Psi_{ML}}$ in Eq. (\ref{eq:maxLogLike}) does not have an analytic solution and therefore must be computed numerically.  Taking the LLR as the cost function in the optimization, each evaluation of $\lambda$, Eq. (\ref{eq:logLike}), requires a comparison of two conditional states, as observed through their respective expectation values of $J_z$.  This, in turn, requires an efficient method for integrating the CSE, Eq. (\ref{eq:CSE}), since a typical minimization algorithm will require many evaluations of $\lambda$.  In general, this is a numerically intensive, as the dimension of the Hilbert space for the collective state in the symmetric Hilbert space grows as $N+1$, and we seek to study the limits for large $N$.  We can substantially reduce this numerical complexity by making an approximation on the measurement-induced dynamics.

To understand the appropriate approximation, let us consider how measurement backaction complicates the description of the dynamics.  In general, the state of the symmetric ensemble of $N$ particles is specified by all distinct symmetrized $K$-body correlation functions of Pauli products, $\expect{\sigma_{\alpha_{1}}^{(1)} \sigma_{\alpha_{2}}^{(2)}\cdots\sigma_{\alpha_{K}}^{(K)}}_{\mathrm{sym}}$, where $\sigma_{\alpha_{i}}^{(i)}$  acts on the $i^{\mathrm{th}}$ spin with $\alpha_{i} \in \{x,y,z\}$ and  $K=1,\cdots,N$.  For the special case of a spin coherent state (SCS), the state is completely specified only by the one-point correlation functions $n_\alpha=\expect{\sigma_\alpha}$ -- the Bloch vector of any of the identical qubits.  The effect of measurement of the collective spin is two-fold: (i) the Bloch vector is stochastically ``kicked'' when conditioned on the noisy measurement record; (ii) higher order correlations (entanglement) are generated between the qubits.  To lowest order, the measurement induced correlations result in spin squeezing~\cite{hammerer_quantum_2010}, specified by two-point correlations.  For stronger measurements all correlations become important.  In the absence of any control, the continuous measurement ultimately becomes projective, yielding a Dicke state (eigenstate of the collective $J_z$) as the steady state of a perfect QND measurement~\cite{stockton_deterministic_2004}.

\begin{figure*}[tbh]
    \begin{center}
        \includegraphics[width=\hsize]{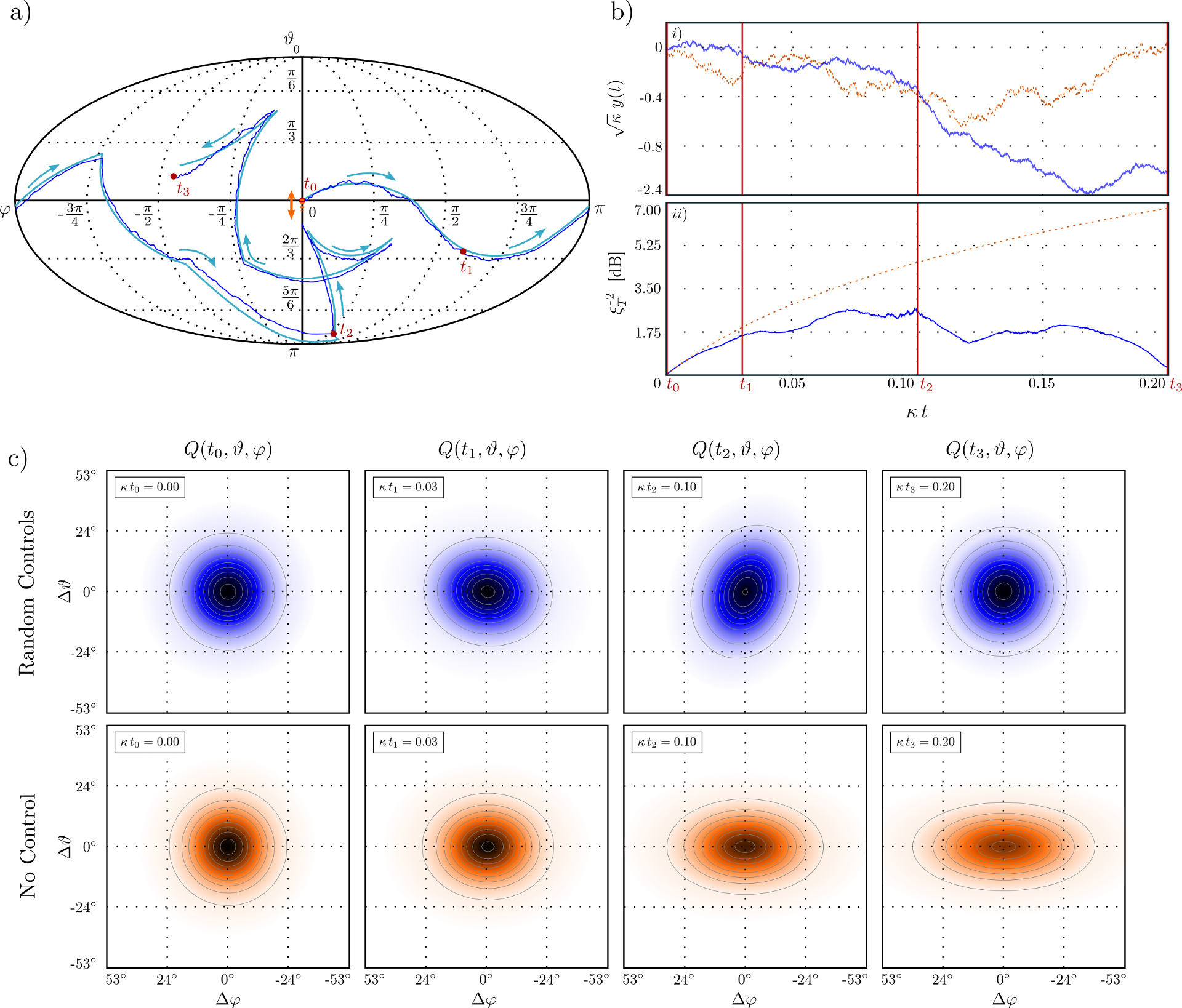}
        \caption{\label{fig:Sqz-Trajectory} (Color online) Simulations for $N = 75$ qubits, initially prepared in a SCS, polarized along $x$.  a) The trajectory of the mean spin on the Bloch sphere for the cases of a randomized control Hamiltonian with measurement (dark blue), a control Hamiltonian without measurement, \emph{i.e.} $\kappa = 0$ (light blue), and no controls but with measurement (dashed orange).  b-\emph{i}) The simulated measurement records in the presence (solid blue) and absence (dotted orange) of a randomized control Hamiltonian.  b-\emph{ii}) The amount of spin squeezing (in dB) generated  the presence (solid blue) and absence (dotted orange) of the controls.  The squeezing is a measure of the correlations between qubits generated by the measurement backaction.  c) Conditional spin $Q$-functions for both trajectories taken at sample times $t_0, t_1, t_2$, and $t_3$.  The random rotations generated by the control Hamiltonian averages out the effect of squeezing, leaving the collective state close to a product SCS.}
	\end{center}
\end{figure*}

In the presence of strong randomized controls, the state evolution is dramatically different.  In addition to causing precession of the mean spin, a transverse magnetic field will generally rotate the reduced uncertainty of the measured component into an orthogonal direction.  The subsequent direction being measured will likely have increased uncertainty, \emph{i.e.}, is \emph{anti-squeezed}.  This measurement will in turn reduce the previously increased uncertainty, resulting in at least a partial cancelation.  The ultimate effect is that by applying rapid rotations about random directions, any spin squeezing produced at early times has a good chance at being undone at later times, and on average, the state remain close to a SCS.

An example of this effect can be seen in Fig. \ref{fig:Sqz-Trajectory}, which contrasts the conditional evolution of a QND measurement of $J_z$  without control to a system subjected to 10 $\pi/2$-rotations about random directions ($\Omega_b = 25\, \pi \,\kappa$) while being continuously measured.  In both simulations we take $N = 75$ qubits ($J = 37.5$) initially prepared in a SCS along $x$.  Figure \ref{fig:Sqz-Trajectory}a shows the trajectory that the mean spin takes on the Bloch sphere under the influence of the controls, both in the absence of a continuous measurement and when conditioned on the solid blue measurement record in Fig. \ref{fig:Sqz-Trajectory}b-\emph{i}.  As a quantitative measure of the two-body correlations present in the system, in Fig. \ref{fig:Sqz-Trajectory}b-\emph{ii} we plot the spin squeezing parameter $\xi_T^2 \equiv \lambda_{\textrm{min}}/J^2$ \cite{toth_spin_2009}, where $\lambda_{\textrm{min}}$ is the minimum eigenvalue of the matrix $G$ with entries
\begin{equation}
G_{ij} = \tfrac{N}{2} \expect{ J_i J_j + J_j J_i}_{\Psi} - (N-1) \expect{ J_i}_{\Psi}\expect{ J_j}_{\Psi}.
\end{equation}
This particular parameter is qualitatively equivalent to the concurrence \cite{yin_spin_2011}, a measure of pair-wise entanglement between qubits.  We see that in the absence of the controls, squeezing grows monotonically, reaching its maximum value at a final time $\kappa t = 0.2$.  However, in the presence of the controls the squeezing does not monotonically increase, but instead reaches a maximum value at time $\kappa t \approx 0.1$, and then returns to a value near zero.

We can see how the controls average out the effect of squeezing and entanglement by plotting the spin-Husimi $Q$-function at various sample times,
\begin{equation}
  Q(t, \vartheta, \varphi) \equiv \frac{N + 1}{4 \pi}\, \abs{\braket{\vartheta, \varphi}{\Psi(t)}}^2,
\end{equation}
where $\ket{\vartheta, \varphi}\equiv \ket{\mathbf{n}(\vartheta, \varphi)}^{\otimes N}$ is a SCS whose Bloch vector $\mathbf{n}$ is parameterized by the spherical coordinate angles $\vartheta$ and $\varphi$.  For spin squeezed states, the $Q$-function takes the form of an approximately Gaussian distribution, centered at the mean spin position and with its minor axis orientated in the squeezing direction.   Figure \ref{fig:Sqz-Trajectory}c shows contour plots the $Q$-function, both with and without controls, at times $\kappa\, t_0 = 0$, $\kappa\, t_1 = 0.03$, $\kappa\,t_2 = 0.1$, and $\kappa\, t_3 = 0.2$.  The $Q$-function in the presence of controls begins as an unsqueezed SCS and proceeds to rotate about $z$ axis, staying roughly near the equator.  During this time it also being squeezed, as its minor axis has strong overlap with the measurement axis.  This continues until $ \kappa t \sim 0.08$, when the controls rotate the mean spin to be near the $-z$ axis.  As it does so, the minor and major axes are also rotated so that by time $\kappa t = 0.1$,  the anti-squeezed major axis is almost aligned with the measurement axis.  The remainder of the evolution returns the mean spin to near the equator, in such a way as to preserve this orientation and subsequently undoes the accumulated squeezing, as seen in the $Q$-function at the final time, $\kappa t_3 = 0.2$.  This is contrasted with the uncontrolled evolution, which shows a sequence of increasingly eccentric ellipses whose minor axes are always orientated along the $z$-axis.

Given these facts, we propose the \emph{ansatz} that the exact conditional state is well approximated by a conditional SCS, a state that is always a separable product, $\ket{\Psi(t)} \approx \ket{\mathbf{n}(t)}^{\otimes N}$, where $\mathbf{n}(t)$ is a conditional single qubit Bloch vector.  This ansatz allows us to extend the continuous measurement QST protocol to include the effect of measurement backaction, by returning an estimate that depends only on the evolution of a single-body density operator.  The direction of the Bloch vector will evolve under the control Hamiltonian, Eq. (\ref{eq:Hcontrol}), with a stochastic component arising from the measurement backaction.  We make this approximation by deriving the conditional evolution $\expect{\V{J}}_{\Psi(t)}$ under the assumption that all moments are computed under a SCS approximation. 
The equation of motion for $\expect{\V{J}}_{\Psi(t)}$ follows from the exact CSE, yielding the It\=o equation,
\begin{widetext}
\begin{equation}
\begin{split}
d\expect{\V{J}} &= \V{b}(t) \times \expect{\V{J}}\, dt -\frac{\kappa}{8}\expect{[J_z,[J_z,\V{J}]]}\, dt  +\frac{\sqrt{\kappa}}{2}\big(\expect{J_z\V{J}+\V{J}J_z} -2\expect{J_z}\expect{\V{J}}\big) \, dv(t) \\
&= \V{b}(t) \times \expect{\V{J}} \, dt -\frac{\kappa}{8}\big(\expect{\V{J}}-\mathbf{e}_z\expect{J_z} \big)\, dt +\frac{\sqrt{\kappa}}{2}\big( \expect{J_z\V{J}+\V{J}J_z}- 2\expect{J_z} \expect{\V{J}}\big)\, dv(t),
\end{split}
\end{equation}
\end{widetext}
where all expectation values are computed under the state $\ket{\Psi(t)}$.  Under the SCS approximation, $\expect{\V{J}} \approx \frac{N}{2} \mathbf{n}(t)$ and $\expect{J_z\V{J}+\V{J}J_z}-2\expect{J_z}\expect{\V{J}} \approx   \frac{N}{2} (\mathbf{e}_z - \expect{\sigma_z}\, \mathbf{n}(t) )$.  The conditional evolution of the Bloch vector $\mathbf{n}(t)$  thus obeys the SDE,
\begin{subequations}\label{eq:SCSBlochVector}
\begin{align}
  d \mathbf{n}(t) &= \left( \V{b}(t) \times \mathbf{n}(t) - \tfrac{1}{8}\kappa\, \left( \mathbf{n}(t)- \mathrm{z}(t)\ \mathbf{e}_z \right) \right) \, dt \\
      &\quad + \half \sqrt{\kappa} \left( \mathbf{e}_z - \mathrm{z}(t)\ \mathbf{n}(t)\right)\, dv(t), \nonumber \\
  d v(t) &=  d y(t) - \sqrt{\kappa} \,  \tfrac{N}{2}\, \mathrm{z}(t) \, dt\label{eq:dvSCS},
\end{align}
\end{subequations}
where $(\mathrm{x, y, z}) =(\expect{\sigma_x}, \expect{\sigma_y},\expect{\sigma_z})$.  This is the same SDE we would derive for the conditional evolution of a single qubit, with the exception that the innovation $v(t)$ expects a signal scaled by the factor $J=N/2$.   Note, this equation is valid for both pure and mixed single qubit states, a fact we exploit in our reconstruction algorithm.

To test the quality of this approximation, we compare the exact evolution of the collective state $\ket{\Psi(t)}$,  governed by the CSE, Eq. (\ref{eq:CSE}), to that given by the SCS approximation, $\ket{\Psi_{SCS}(t)}=\ket{\mathbf{n}(t)}^{\otimes N}$, governed by Eq. (\ref{eq:SCSBlochVector}).  Given the same SCS initial condition in both cases, we compare these two states in two different ways.  Firstly, we compute the fidelity between these two states, $\mathcal{F} = \abs{\braket{\Psi_{SCS}(t)}{\Psi(t)}}^2$ as a function of time, and secondly, we compute the RMS error between $\expect{J_z}_{\Psi(t)}$ and $\expect{J_z}_{\Psi_{SCS}(t)}$ as defined by the quantity
\begin{equation}\label{eq:RMSZError}
  \Delta \mathrm{z _{err}}(t) \equiv \sqrt{ \expect{\left(\frac{1}{J} \expect{J_z}_{\Psi(t)} - \mathrm{z}(t)\right)^2 }_\nu }.
\end{equation}
The expectation values have been scaled by the total spin length $J$ to allow for a comparison between different values of $N$.  This quantity impacts the performance of the estimator, since any error in $\expect{J_z}$ directly impacts the LLR.  The ensemble average is computed for $\nu = 100$ unit vectors uniformly sampled over the Bloch sphere, and use only a single noise realization per state.

\begin{figure}[thb]
	\begin{center}
		\includegraphics[width=1\hsize]{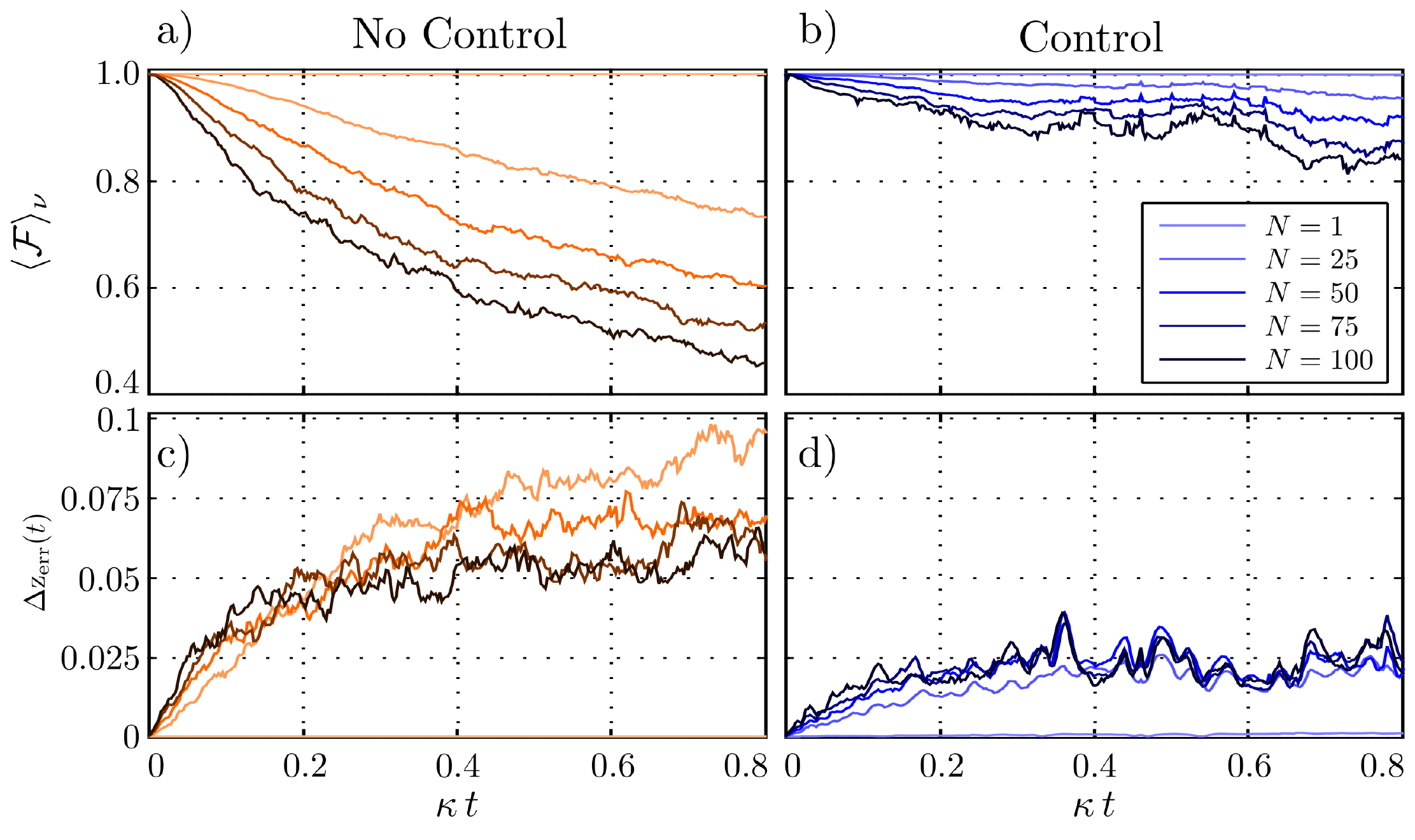}
   		 \caption{\label{fig:SCSApproxFidelity} (Color online) Performance of the separable SCS approximation.  Average fidelity between the exact state $\ket{\Psi(t)}$ and the SCS $\ket{\mathbf{n}(t)}^{\otimes N}$ as a function of time for a) no controls and b) applying 40 $\pi/2$-rotations over a time $\kappa T = 0.8$.  RMS error $\Delta \mathrm{z _{err}}(t)$ as defined in Eq. (\ref{eq:RMSZError}) is plotted for c) no controls and d) with the same control law as in b).  The average is over $\nu = 100$ random initial unit vectors, uniformly sampled over the Bloch sphere, with a single noise realization per state.  We make these comparisons for $N = 1, 25, 50, 75, 100$ qubits. plotted with a correspondingly increasing contrast and decreasing fidelity.}
	\end{center}
\end{figure}

Figure \ref{fig:SCSApproxFidelity} shows this average fidelity, $\expect{\mathcal{F}}_\nu$,  for a variety of numbers of qubits, $N$, both with and without 40 $\pi/2$-rotations about random directions, for a total measurement time $\kappa\, T = 0.8$.  The SCS approximation performs poorly in the absence of the controls and for large $N$, showing a worst case average fidelity of $\expect{\mathcal{F}}_\nu \sim 0.47$ for $N = 100$.  In the presence of the controls the approximation performs well, maintaining the fidelity at a level $\expect{\mathcal{F}}_\nu > 0.80$ for all $N$ tested.  The non-monotonic decrease in the average fidelity implies that the controls could be optimized to maximize this value, however, it is unclear if such an optimization would return an optimal tomographic estimate. Figures \ref{fig:SCSApproxFidelity}c and \ref{fig:SCSApproxFidelity}d show that for all of the $N$ that we simulated, the SCS approximation tracks the mean spin with $ \Delta \mathrm{z _{err}}(t) < 0.1$, and that in the presence of the controls, $ \Delta \mathrm{z _{err}}(t) \sim 0.025$.  The case $N=1$ shows that Eq. (\ref{eq:SCSBlochVector}) exactly reproduces the single qubit evolution, up to numerical precision.

\section{Numeric Simulations\label{sec:Results}}

Armed with the SCS approximation we are able to efficiently compute an approximate version of the LLR given in Eq. (\ref{eq:logLike}).  We now test the performance of our tomographic procedure via a series of numerical simulations.  In the absence of a closed form solution to Eqs. (\ref{eq:logLike}) and (\ref{eq:SCSBlochVector}), we must also find $\ket{\Psi_{ML}}$ through a numerical search.    While we may choose from any number of algorithms (\emph{e.g.} gradient assent) we use a particularly simple procedure here.   As the Bloch sphere is such a small search space, we simply sample a suitably dense set of initial conditions and then choose as our estimate the element that maximizes $\lambda$.  We operate with a density of samples such that the average infidelity between nearest neighbors is $\sim 6\times 10^{-4}$.  This ensures that we will obtain an estimate that is sufficiently close to the true state.  As an example, given $N = 100$ qubits, the optimum POVM bound sets an average infidelity of $0.01$ \cite{massar_optimal_1995}, implying that any deficits in our procedure should not be attributed to the finite number of samples.

In practice, we need to consider an additional step in our protocol.  A CSE with an informationally complete measurement record is in principle stable~\cite{van_handel_stability_2009}.  This means that given a measurement record generated from an initial state $\ket{\Psi(0)}$, it is possible to integrate a CSE from any initial condition $\ket{\Psi(0)'} \ne \ket{\Psi(0)}$, such that $\ket{\Psi(t)'} \rightarrow \ket{\Psi(t)}$ as $t \rightarrow \infty$.  In a sense, this means that the CSE is self-correcting for the initial misinformation.    Unfortunately, we find that the numerical stability of both Eqs. (\ref{eq:CSE}) and (\ref{eq:SCSBlochVector}) is quite poor when the initial condition is nearly orthogonal to the true state.  This affects our reconstruction procedure because an instability in computing either the candidate or reference state can result in $\lambda$ reaching arbitrarily large or small values.  To correct for this issue, we first compute $\lambda$ for \emph{mixed} initial conditions, and then for a spread of pure states in the direction of the most likely mixed state.  This two-step procedure greatly improves the numerical stability because a mixed state polarized in a direction orthogonal to the true state still has some overlap with that state.

\begin{figure}[thb]
    \begin{center}
        \includegraphics[width=\hsize]{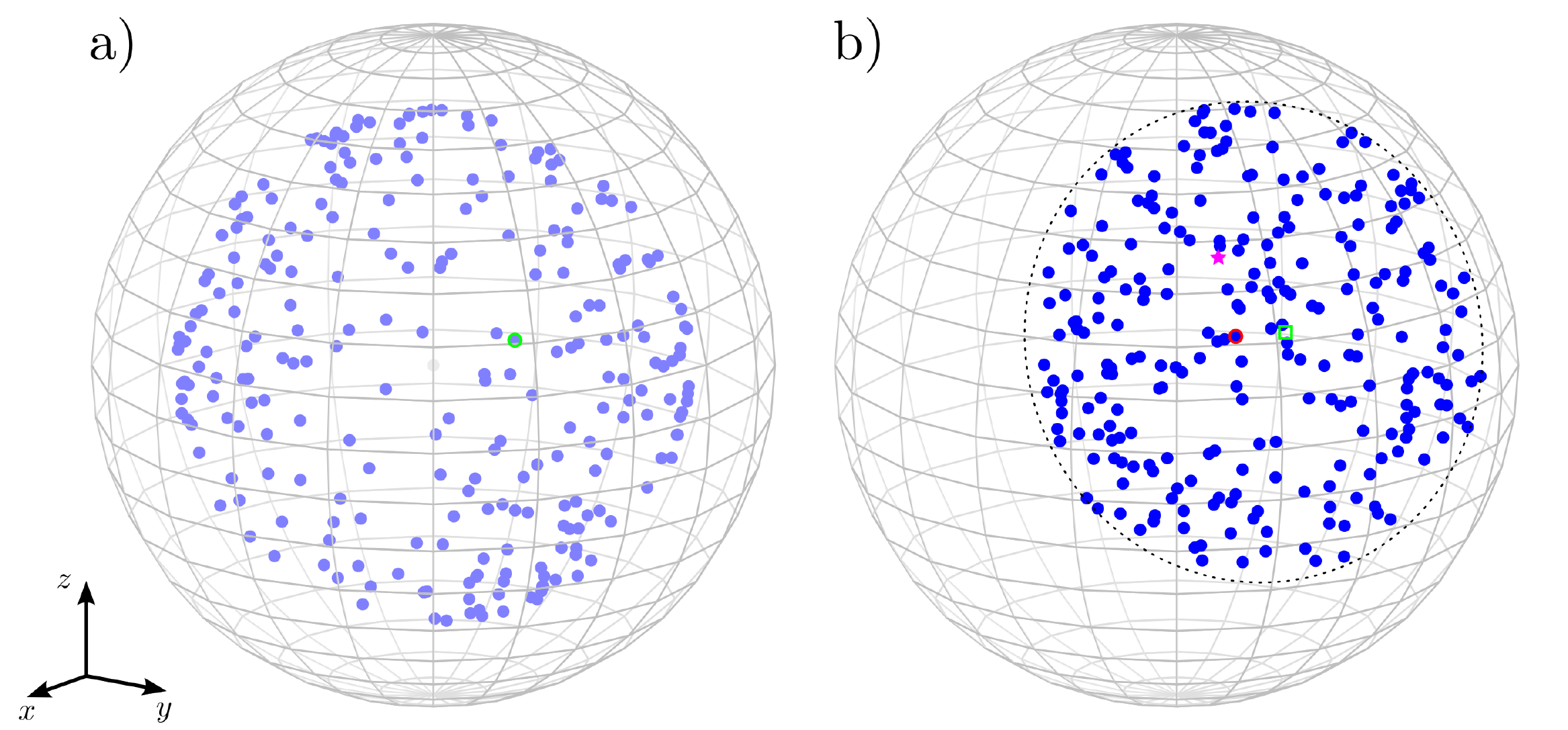}
        \caption{ \label{fig:SamplePts} (Color online) Typical Sample Distributions.
         a) The set $\mathcal{N}_1$ shown inside the unit sphere for $M_1 = 250$ and $\norm{\mathbf{n}_m(0)} = 3/4$.  The point that maximized $\lambda_{scs}$ relative $\mathbf{n}_r(0) = 0$, given a measurement record made with $N = 75$ qubits, is circled in green.  b) The resample set $\mathcal{N}_2$  for $M_2 = 250$ is plotted in blue on the Bloch sphere, with the maximum angular deviation of $\pi/4$ indicated by the dashed line.  Also shown is the reference point $\mathbf{n}_r'$ (green square) and the true initial state (magenta star).  The resample point that maximizes $\lambda_{scs}$ is circled in red.}
	\end{center}
\end{figure}

In the first step, we use $M_1 = 250$ isotropically distributed mixed states, whose Bloch vectors form the set $\mathcal{N}_1 = \big\{ \mathbf{n}_m \in \mathbbm{R}^3 :\ \| \mathbf{n}_m \| = 3/4,\  m = 1,\cdots, M_1\big\}$.  To identify an acceptable pure reference state, we find the mixed state sample $\mathbf{n}^{\star} \in \mathcal{N}_1$ that maximizes the approximate LLR,
\begin{multline}\label{eq:logLikeSCS}
  \lambda_{scs}(\mathbf{n}_m,\mathbf{n}_r ) = \tfrac{\sqrt{\kappa}\, N}{2} \int^{T}_{0} [ \mathrm{z}_m(s) - \mathrm{z}_r(s)]\, dy(s) \\
        -\tfrac{\kappa\, N^2 }{8} \int_0^T \left[  \mathrm{z}_m(s)^2 - \mathrm{z}_r(s)^2\, \right]\, ds,
\end{multline}
where we choose the unbiased reference initial condition $\| \mathbf{n}_r(0)\| = 0$.  From this mixed state, we then define the new reference vector $\mathbf{n}'_r \equiv \mathbf{n}^{\star}/ \|\mathbf{n}^{\star}\|$ and uniformly sample $M_2 = 250$ pure states within a neighborhood of this vector.  Specifically, we form the set
$\mathcal{N}_2 =  \Big\{\mathbf{n}_m \in \mathbbm{R}^3 : \| \mathbf{n}_m \| = 1,\, \mathbf{n}_m \cdot \mathbf{n}_{r}'  \ge \cos( \pi/4 ), \ m = 1,\cdots, M_2\Big\}$.  We then report as an estimate, the single qubit state $\ket{\mathbf{n}_{ML}}$ whose Bloch vector $\mathbf{n}_{ML} \in \mathcal{N}_2$ maximizes $\lambda_{scs}(\mathbf{n}_m, \mathbf{n}'_r)$. Figure \ref{fig:SamplePts} shows a typical realization of both sample sets, for a simulation over $N = 75$ qubits.

In order to characterize the performance of our protocol, we perform a series of numerical simulations for a variety of $N$.  In each simulation we wish to compare the average infidelity between our estimate and the true input state, $1 - \expect{\mathcal{F}}_\nu$, averaged over uniformly sampled inputs and measurement realizations.  We will also compare our protocol to two different measurement schemes.  The first is to the fundamental bound set by the optimum POVM, with $1 - \expect{\mathcal{F}}_{\rm{opt}} = 1/(N+2)$~\cite{massar_optimal_1995}.  The second comparison is to an alternative model of the continuous measurement, one that completely ignores measurement backaction.  In other words, we wish to compare the above model to a model where the measurement record is approximated by
\begin{equation}\label{eq:ytBackactionFree}
    \tilde{y}(t) \approx w(t) + \tfrac{  \sqrt{\kappa}\, N}{2} \int_0^t  \braOket{\mathbf{n}(0)}{\sigma_z(s)}{\mathbf{n}(0)}\, ds
\end{equation}
where $\sigma_z(s)$ is the Heisenberg evolved Pauli-z operator and $w(t)$ is a Wiener process.  This model is equivalent to the $\gamma_{diss} = 0$ limit of the single atom density matrix $\tilde{\rho}$ defined in Eq. (\ref{eq:singleAtomMasterEq}).  While such a model is a good approximation when the total measurement time is very short compared to $1/\kappa$, we expect the effect of measurement backaction to have a significant impact on our estimator.

To make a fair comparison, we use a nearly identical algorithm in the backaction-free case as in the estimator described above.  In this case we no longer have a problem with the numerical stability of our estimator because the Heisenberg equation of motion for $\sigma_z$ is independent of the state.  Therefore, we need not perform a two-step sampling procedure.  We thus uniformly sample $M$ pure Bloch vectors with a density equal to the final density of samples that we used in the procedure above, which requires $M = 1700$.  We then choose the sampled state that maximizes a backaction-free version of the LLR, where the conditional expectation values $\expect{J_z}_{\Psi_i(t)}$ are replaced by $N\, \langle \mathbf{n}_i|\sigma_z(t)|\mathbf{n}_i\rangle/2$, with the first sample $\mathbf{n}_1$ serving as the reference.

\begin{figure}[htb]
    \begin{center}
        \includegraphics[width=\hsize]{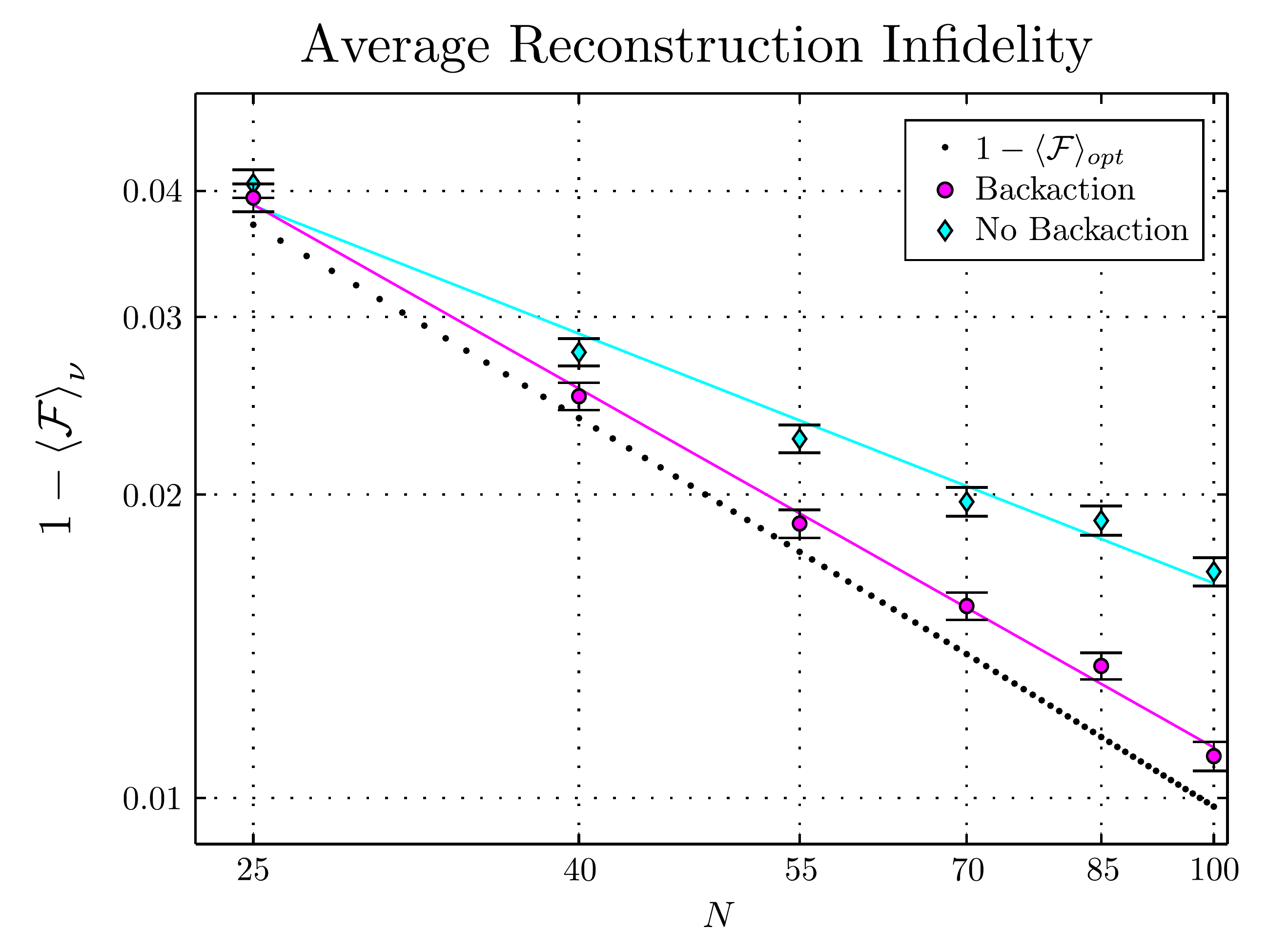}
        \caption{ \label{fig:FidelPlot}  (Color online) Average quantum state reconstruction infidelities (log-log axes) for different estimators.    Circles show the performance of the MLE based on the LLR, Eq. (\ref{eq:maxLogLike}), with the SCS approximation to the dynamical evolution, Eq. (\ref{eq:logLikeSCS}).  A power law fit to this data yields $1-\expect{\mathcal{F}}_\nu\propto N^{-0.89}$, which is close to the bound set by the optimal POVM, $1 - \expect{\mathcal{F}}_\nu \approx N^{-1}$ (dotted line).   Diamonds show the performance of an estimator that assumes a backaction-free a measurement model given by Eq. (\ref{eq:ytBackactionFree}) and achieves a power law scaling, $1-\expect{\mathcal{F}}_\nu \propto N^{-0.60}$.    Error bars show a standard error of $\pm \sqrt{\text{Var}[1-\mathcal{F}]/\nu}$.}
	\end{center}
\end{figure}

Figure \ref{fig:FidelPlot} shows the results of numerical simulations for our reconstruction procedure with and without backaction.  Plotted on a log-log scale is the average infidelity, $1 - \expect{\mathcal{F}}_\nu$, for $N = 25, 40, 55, 70, 85, 100$ qubits.  For every $N$, we average over $\nu = 1000$ initial single qubit states with a single measurement realization per state.  Every simulation used the same control law, with 40 randomized $\pi/2$ rotations, as well as a final time $\kappa\, T = 0.8$.  Also shown are linear-least-squares fits to a power law, $a N^b$.  With backaction, the best fit parameters are $a = 0.69 \pm 0.08$ and $b = -0.89 \pm 0.03$, and without backaction $a = 0.29 \pm 0.06$ and $b = -0.62 \pm 0.05$. These exponents are compared to the optimal scaling of $1/N$.  By implementing the SCS approximation, we have introduced $\lesssim 5\%$ systematic errors in computing $\expect{J_z}$, which propagates into the approximate LLR, $\lambda_{scs}$, ultimately contributing to the suboptimal scaling.

The performance of the backaction-free estimator is best understood by considering not only final state reconstruction given the entire measurement record, but also the family of estimates generated by using data up a time  $0 \le t \le T$.  The stability of the CSE implies that for initial conditions $\Psi_1(0) \ne \Psi_2(0)$ we have the convergence $\expect{J_z}_{\Psi_1(t)} - \expect{J_z}_{\Psi_2(t)} \rightarrow 0$ as $t \rightarrow \infty$.  The effect of this is that the LLR will either reach or asymptotically approach a steady-state value at long times.  This convergence is necessarily implemented through the innovation, which occurs faster for larger $N$, as follows from Eq. (\ref{eq:dvSCS}).  However, the unitary evolution in the backaction-free scheme is unable to implement such a convergence and thus, the LLR will never reach steady-state.  This ultimately biases the estimate away from the true state at long times, which can been seen in the poor performance of the backaction-free estimate for large $N$.

\section{Summary and Outlook}
We have studied a protocol that performs quantum state tomography using a single continuous measurement record of an ensemble of identical copies when the system is subjected to dynamical control and measurement backaction.  We have considered the simplest case -- estimation of the direction of the Bloch vector of a pure qubit in the absence of decoherence and systematic errors.  This allowed us to focus on the effects of measurement backaction that complicate the estimator due to the nonlinearity of the conditional state evolution, and the many-body nature of the dynamics induced by the entangling QND measurement.

We formulated a maximum-likelihood estimator, and showed that it is possible to obtain a high-fidelity reconstruction an initial SCS using only a single realization of a continuous collective measurement and dynamical control.  Numerical simulations indicate that this estimate nearly reaches the bound set by the optimal POVM.  By failing to include the effect of measurement backaction in the conditional dynamics of the mean spin direction, an otherwise equivalent estimator becomes biased towards a poorer estimate at long times.

A key feature of our estimator was a simplification of the dynamical model in which the effects of measurement backaction act solely to induce random kicks on the direction of the Bloch vector while the entangling effects of the QND measure are small.  This followed from the fact that the random rotations of the Bloch vector generated by the control Hamiltonian acted to average out the effects of squeezing and as well as any higher-order correlations between the qubits.   A next natural extension is to include two-point atom-atom correlations in our estimator.  This could allow us to improve the fidelity of our estimator and generalize the class of states we can reconstruct, including spin squeezed states or  other Gaussian states described by two-atom correlations.

Going beyond two-body correlations, the matrix product state formalism provides a natural framework for studying many-body effects~\cite{perez-garcia_matrix_2007, verstraete_matrix_2008}. The application of the matrix product state ansatz to maximum likelihood tomography has been studied~\cite{baumgratz_scalable_2013}, with good results. By translating this to the continuous measurement setting discussed here, one might be able to extract many-body correlations efficiently and robustly, of particular interest in the context of quantum simulators~\cite{Wright_QuantumSimulation_2012}.

\vspace{.25 in}

\emph{Acknowledgments} We gratefully acknowledge Josh Combes, Chris Ferrie, and Ben Baragiola for helpful discussions.  RLC and IHD were supported by NSF Grants PHY-1307520 and PHY-1212445.  CAR thanks the EU (SIQS, RAQUEL, COST) and the BMBF for their support.

\bibliography{qubitClean}

\begin{thebibliography}{26}
\expandafter\ifx\csname natexlab\endcsname\relax\def\natexlab#1{#1}\fi
\expandafter\ifx\csname bibnamefont\endcsname\relax
  \def\bibnamefont#1{#1}\fi
\expandafter\ifx\csname bibfnamefont\endcsname\relax
  \def\bibfnamefont#1{#1}\fi
\expandafter\ifx\csname citenamefont\endcsname\relax
  \def\citenamefont#1{#1}\fi
\expandafter\ifx\csname url\endcsname\relax
  \def\url#1{\texttt{#1}}\fi
\expandafter\ifx\csname urlprefix\endcsname\relax\def\urlprefix{URL }\fi
\providecommand{\bibinfo}[2]{#2}
\providecommand{\eprint}[2][]{\url{#2}}

\bibitem[{\citenamefont{Haffner et~al.}(2005)\citenamefont{Haffner, Hansel,
  Roos, Benhelm, Chek{--}al{--}kar, Chwalla, Korber, Rapol, Riebe, Schmidt
  et~al.}}]{haffner_scalable_2005}
\bibinfo{author}{\bibfnamefont{H.}~\bibnamefont{Haffner}},
  \bibinfo{author}{\bibfnamefont{W.}~\bibnamefont{Hansel}},
  \bibinfo{author}{\bibfnamefont{C.~F.} \bibnamefont{Roos}},
  \bibinfo{author}{\bibfnamefont{J.}~\bibnamefont{Benhelm}},
  \bibinfo{author}{\bibfnamefont{D.}~\bibnamefont{Chek{--}al{--}kar}},
  \bibinfo{author}{\bibfnamefont{M.}~\bibnamefont{Chwalla}},
  \bibinfo{author}{\bibfnamefont{T.}~\bibnamefont{Korber}},
  \bibinfo{author}{\bibfnamefont{U.~D.} \bibnamefont{Rapol}},
  \bibinfo{author}{\bibfnamefont{M.}~\bibnamefont{Riebe}},
  \bibinfo{author}{\bibfnamefont{P.~O.} \bibnamefont{Schmidt}},
  \bibnamefont{et~al.}, \bibinfo{journal}{Nature}
  \textbf{\bibinfo{volume}{438}}, \bibinfo{pages}{643} (\bibinfo{year}{2005}).

\bibitem[{\citenamefont{Leibfried et~al.}(2005)\citenamefont{Leibfried, Knill,
  Seidelin, Britton, Blakestad, Chiaverini, Hume, Itano, Jost, Langer
  et~al.}}]{leibfried_creation_2005}
\bibinfo{author}{\bibfnamefont{D.}~\bibnamefont{Leibfried}},
  \bibinfo{author}{\bibfnamefont{E.}~\bibnamefont{Knill}},
  \bibinfo{author}{\bibfnamefont{S.}~\bibnamefont{Seidelin}},
  \bibinfo{author}{\bibfnamefont{J.}~\bibnamefont{Britton}},
  \bibinfo{author}{\bibfnamefont{R.~B.} \bibnamefont{Blakestad}},
  \bibinfo{author}{\bibfnamefont{J.}~\bibnamefont{Chiaverini}},
  \bibinfo{author}{\bibfnamefont{D.~B.} \bibnamefont{Hume}},
  \bibinfo{author}{\bibfnamefont{W.~M.} \bibnamefont{Itano}},
  \bibinfo{author}{\bibfnamefont{J.~D.} \bibnamefont{Jost}},
  \bibinfo{author}{\bibfnamefont{C.}~\bibnamefont{Langer}},
  \bibnamefont{et~al.}, \bibinfo{journal}{Nature}
  \textbf{\bibinfo{volume}{438}}, \bibinfo{pages}{639} (\bibinfo{year}{2005}).

\bibitem[{\citenamefont{Silberfarb et~al.}(2005)\citenamefont{Silberfarb,
  Jessen, and Deutsch}}]{silberfarb_quantum_2005}
\bibinfo{author}{\bibfnamefont{A.}~\bibnamefont{Silberfarb}},
  \bibinfo{author}{\bibfnamefont{P.~S.} \bibnamefont{Jessen}},
  \bibnamefont{and} \bibinfo{author}{\bibfnamefont{I.~H.}
  \bibnamefont{Deutsch}}, \bibinfo{journal}{Phys. Rev. Lett.}
  \textbf{\bibinfo{volume}{95}}, \bibinfo{pages}{030402}
  (\bibinfo{year}{2005}).

\bibitem[{\citenamefont{Riofr\'io et~al.}(2011)\citenamefont{Riofr\'io, Jessen,
  and Deutsch}}]{riofrio_quantum_2011}
\bibinfo{author}{\bibfnamefont{C.~A.} \bibnamefont{Riofr\'io}},
  \bibinfo{author}{\bibfnamefont{P.~S.} \bibnamefont{Jessen}},
  \bibnamefont{and} \bibinfo{author}{\bibfnamefont{I.~H.}
  \bibnamefont{Deutsch}}, \bibinfo{journal}{J. Phys. B}
  \textbf{\bibinfo{volume}{44}}, \bibinfo{pages}{154007}
  (\bibinfo{year}{2011}).

\bibitem[{\citenamefont{Smith et~al.}(2006)\citenamefont{Smith, Silberfarb,
  Deutsch, and Jessen}}]{smith_efficient_2006}
\bibinfo{author}{\bibfnamefont{G.~A.} \bibnamefont{Smith}},
  \bibinfo{author}{\bibfnamefont{A.}~\bibnamefont{Silberfarb}},
  \bibinfo{author}{\bibfnamefont{I.~H.} \bibnamefont{Deutsch}},
  \bibnamefont{and} \bibinfo{author}{\bibfnamefont{P.~S.}
  \bibnamefont{Jessen}}, \bibinfo{journal}{Phys. Rev. Lett.}
  \textbf{\bibinfo{volume}{97}}, \bibinfo{pages}{180403}
  (\bibinfo{year}{2006}).

\bibitem[{\citenamefont{Smith et~al.}(2013)\citenamefont{Smith, Riofr\'io,
  Anderson, Sosa-Martinez, Deutsch, and Jessen}}]{smith_quantum_2013}
\bibinfo{author}{\bibfnamefont{A.}~\bibnamefont{Smith}},
  \bibinfo{author}{\bibfnamefont{C.~A.} \bibnamefont{Riofr\'io}},
  \bibinfo{author}{\bibfnamefont{B.~E.} \bibnamefont{Anderson}},
  \bibinfo{author}{\bibfnamefont{H.}~\bibnamefont{Sosa-Martinez}},
  \bibinfo{author}{\bibfnamefont{I.~H.} \bibnamefont{Deutsch}},
  \bibnamefont{and} \bibinfo{author}{\bibfnamefont{P.~S.}
  \bibnamefont{Jessen}}, \bibinfo{journal}{Phys. Rev. A}
  \textbf{\bibinfo{volume}{87}}, \bibinfo{pages}{030102}
  (\bibinfo{year}{2013}).

\bibitem[{\citenamefont{Chaudhury et~al.}(2007)\citenamefont{Chaudhury, Merkel,
  Herr, Silberfarb, Deutsch, and Jessen}}]{chaudhury_quantum_2007}
\bibinfo{author}{\bibfnamefont{S.}~\bibnamefont{Chaudhury}},
  \bibinfo{author}{\bibfnamefont{S.}~\bibnamefont{Merkel}},
  \bibinfo{author}{\bibfnamefont{T.}~\bibnamefont{Herr}},
  \bibinfo{author}{\bibfnamefont{A.}~\bibnamefont{Silberfarb}},
  \bibinfo{author}{\bibfnamefont{I.~H.} \bibnamefont{Deutsch}},
  \bibnamefont{and} \bibinfo{author}{\bibfnamefont{P.~S.}
  \bibnamefont{Jessen}}, \bibinfo{journal}{Phys. Rev. Lett.}
  \textbf{\bibinfo{volume}{99}}, \bibinfo{pages}{163002}
  (\bibinfo{year}{2007}).

\bibitem[{\citenamefont{Merkel et~al.}(2008)\citenamefont{Merkel, Jessen, and
  Deutsch}}]{merkel_quantum_2008}
\bibinfo{author}{\bibfnamefont{S.~T.} \bibnamefont{Merkel}},
  \bibinfo{author}{\bibfnamefont{P.~S.} \bibnamefont{Jessen}},
  \bibnamefont{and} \bibinfo{author}{\bibfnamefont{I.~H.}
  \bibnamefont{Deutsch}}, \bibinfo{journal}{Phys. Rev. A}
  \textbf{\bibinfo{volume}{78}}, \bibinfo{pages}{023404}
  (\bibinfo{year}{2008}).

\bibitem[{\citenamefont{Smith et~al.}(2004)\citenamefont{Smith, Chaudhury,
  Silberfarb, Deutsch, and Jessen}}]{smith_continuous_2004}
\bibinfo{author}{\bibfnamefont{G.~A.} \bibnamefont{Smith}},
  \bibinfo{author}{\bibfnamefont{S.}~\bibnamefont{Chaudhury}},
  \bibinfo{author}{\bibfnamefont{A.}~\bibnamefont{Silberfarb}},
  \bibinfo{author}{\bibfnamefont{I.~H.} \bibnamefont{Deutsch}},
  \bibnamefont{and} \bibinfo{author}{\bibfnamefont{P.~S.}
  \bibnamefont{Jessen}}, \bibinfo{journal}{Phys. Rev. Lett.}
  \textbf{\bibinfo{volume}{93}}, \bibinfo{pages}{163602}
  (\bibinfo{year}{2004}).

\bibitem[{\citenamefont{Hammerer et~al.}(2010)\citenamefont{Hammerer,
  S{\o}rensen, and Polzik}}]{hammerer_quantum_2010}
\bibinfo{author}{\bibfnamefont{K.}~\bibnamefont{Hammerer}},
  \bibinfo{author}{\bibfnamefont{A.~S.} \bibnamefont{S{\o}rensen}},
  \bibnamefont{and} \bibinfo{author}{\bibfnamefont{E.~S.}
  \bibnamefont{Polzik}}, \bibinfo{journal}{Rev. Mod. Phys.}
  \textbf{\bibinfo{volume}{82}}, \bibinfo{pages}{1041–1093}
  (\bibinfo{year}{2010}).

\bibitem[{\citenamefont{Massar and Popescu}(1995)}]{massar_optimal_1995}
\bibinfo{author}{\bibfnamefont{S.}~\bibnamefont{Massar}} \bibnamefont{and}
  \bibinfo{author}{\bibfnamefont{S.}~\bibnamefont{Popescu}},
  \bibinfo{journal}{Phys. Rev. Lett.} \textbf{\bibinfo{volume}{74}},
  \bibinfo{pages}{1259} (\bibinfo{year}{1995}).

\bibitem[{\citenamefont{Bagan et~al.}(2005)\citenamefont{Bagan, Monras, and
  Mu\~noz{--}Tapia}}]{bagan_comprehensive_2005}
\bibinfo{author}{\bibfnamefont{E.}~\bibnamefont{Bagan}},
  \bibinfo{author}{\bibfnamefont{A.}~\bibnamefont{Monras}}, \bibnamefont{and}
  \bibinfo{author}{\bibfnamefont{R.}~\bibnamefont{Mu\~noz{--}Tapia}},
  \bibinfo{journal}{Phys. Rev. A} \textbf{\bibinfo{volume}{71}},
  \bibinfo{pages}{062318} (\bibinfo{year}{2005}).

\bibitem[{\citenamefont{Deutsch and Jessen}(2010)}]{deutsch_quantum_2010}
\bibinfo{author}{\bibfnamefont{I.~H.} \bibnamefont{Deutsch}} \bibnamefont{and}
  \bibinfo{author}{\bibfnamefont{P.~S.} \bibnamefont{Jessen}},
  \bibinfo{journal}{Opt. Commun.} \textbf{\bibinfo{volume}{283}},
  \bibinfo{pages}{681} (\bibinfo{year}{2010}).

\bibitem[{\citenamefont{Baragiola et~al.}(2014)\citenamefont{Baragiola, Norris,
  Monta\~{n}o, Mickelson, Jessen, and
  Deutsch}}]{baragiola_three-dimensional_2014}
\bibinfo{author}{\bibfnamefont{B.~Q.} \bibnamefont{Baragiola}},
  \bibinfo{author}{\bibfnamefont{L.~M.} \bibnamefont{Norris}},
  \bibinfo{author}{\bibfnamefont{E.}~\bibnamefont{Monta\~{n}o}},
  \bibinfo{author}{\bibfnamefont{P.~G.} \bibnamefont{Mickelson}},
  \bibinfo{author}{\bibfnamefont{P.~S.} \bibnamefont{Jessen}},
  \bibnamefont{and} \bibinfo{author}{\bibfnamefont{I.~H.}
  \bibnamefont{Deutsch}}, \bibinfo{journal}{Phys. Rev. A}
  \textbf{\bibinfo{volume}{89}}, \bibinfo{pages}{033850}
  (\bibinfo{year}{2014}).

\bibitem[{\citenamefont{Bouten et~al.}(2007)\citenamefont{Bouten, van Handel,
  and James}}]{bouten_introduction_2007}
\bibinfo{author}{\bibfnamefont{L.}~\bibnamefont{Bouten}},
  \bibinfo{author}{\bibfnamefont{R.}~\bibnamefont{van Handel}},
  \bibnamefont{and} \bibinfo{author}{\bibfnamefont{M.~R.} \bibnamefont{James}},
  \bibinfo{journal}{{SIAM} J. Control Optim.} \textbf{\bibinfo{volume}{46}},
  \bibinfo{pages}{2199} (\bibinfo{year}{2007}).

\bibitem[{\citenamefont{Jacobs and Steck}(2006)}]{jacobs_straightforward_2006}
\bibinfo{author}{\bibfnamefont{K.}~\bibnamefont{Jacobs}} \bibnamefont{and}
  \bibinfo{author}{\bibfnamefont{D.~A.} \bibnamefont{Steck}},
  \bibinfo{journal}{Contemp. Phys.} \textbf{\bibinfo{volume}{47}},
  \bibinfo{pages}{279} (\bibinfo{year}{2006}).

\bibitem[{\citenamefont{Wiseman and Milburn}(2010)}]{wiseman_quantum_2010}
\bibinfo{author}{\bibfnamefont{H.~M.} \bibnamefont{Wiseman}} \bibnamefont{and}
  \bibinfo{author}{\bibfnamefont{G.~J.} \bibnamefont{Milburn}},
  \emph{\bibinfo{title}{Quantum Measurement and Control}}
  (\bibinfo{publisher}{Cambridge University Press}, \bibinfo{year}{2010}).

\bibitem[{\citenamefont{Liptser and Shiriaev}(2001)}]{liptser_statistics_2001}
\bibinfo{author}{\bibfnamefont{R.~S.} \bibnamefont{Liptser}} \bibnamefont{and}
  \bibinfo{author}{\bibfnamefont{A.~N.} \bibnamefont{Shiriaev}},
  \emph{\bibinfo{title}{Statistics of Random Processes: I. General Theory}}
  (\bibinfo{publisher}{Springer}, \bibinfo{year}{2001}).

\bibitem[{\citenamefont{Stockton et~al.}(2004)\citenamefont{Stockton, van
  Handel, and Mabuchi}}]{stockton_deterministic_2004}
\bibinfo{author}{\bibfnamefont{J.~K.} \bibnamefont{Stockton}},
  \bibinfo{author}{\bibfnamefont{R.}~\bibnamefont{van Handel}},
  \bibnamefont{and} \bibinfo{author}{\bibfnamefont{H.}~\bibnamefont{Mabuchi}},
  \bibinfo{journal}{Phys. Rev. A} \textbf{\bibinfo{volume}{70}},
  \bibinfo{pages}{022106} (\bibinfo{year}{2004}).

\bibitem[{\citenamefont{T\'oth et~al.}(2009)\citenamefont{T\'oth, Knapp,
  G\"uhne, and Briegel}}]{toth_spin_2009}
\bibinfo{author}{\bibfnamefont{G.}~\bibnamefont{T\'oth}},
  \bibinfo{author}{\bibfnamefont{C.}~\bibnamefont{Knapp}},
  \bibinfo{author}{\bibfnamefont{O.}~\bibnamefont{G\"uhne}}, \bibnamefont{and}
  \bibinfo{author}{\bibfnamefont{H.~J.} \bibnamefont{Briegel}},
  \bibinfo{journal}{Phys. Rev. A} \textbf{\bibinfo{volume}{79}},
  \bibinfo{pages}{042334} (\bibinfo{year}{2009}).

\bibitem[{\citenamefont{Yin et~al.}(2011)\citenamefont{Yin, Wang, Ma, and
  Wang}}]{yin_spin_2011}
\bibinfo{author}{\bibfnamefont{X.}~\bibnamefont{Yin}},
  \bibinfo{author}{\bibfnamefont{X.}~\bibnamefont{Wang}},
  \bibinfo{author}{\bibfnamefont{J.}~\bibnamefont{Ma}}, \bibnamefont{and}
  \bibinfo{author}{\bibfnamefont{X.}~\bibnamefont{Wang}}, \bibinfo{journal}{J
  Phys. B} \textbf{\bibinfo{volume}{44}}, \bibinfo{pages}{015501}
  (\bibinfo{year}{2011}).

\bibitem[{\citenamefont{Van~Handel}(2009)}]{van_handel_stability_2009}
\bibinfo{author}{\bibfnamefont{R.}~\bibnamefont{Van~Handel}},
  \bibinfo{journal}{Infin. Dimens. Annal. Qu.} \textbf{\bibinfo{volume}{12}},
  \bibinfo{pages}{153} (\bibinfo{year}{2009}).

\bibitem[{\citenamefont{Perez{--}Garcia
  et~al.}(2007)\citenamefont{Perez{--}Garcia, Verstraete, Wolf, and
  Cirac}}]{perez-garcia_matrix_2007}
\bibinfo{author}{\bibfnamefont{D.}~\bibnamefont{Perez{--}Garcia}},
  \bibinfo{author}{\bibfnamefont{F.}~\bibnamefont{Verstraete}},
  \bibinfo{author}{\bibfnamefont{M.~M.} \bibnamefont{Wolf}}, \bibnamefont{and}
  \bibinfo{author}{\bibfnamefont{J.~I.} \bibnamefont{Cirac}},
  \bibinfo{journal}{Quantum Info. Comput.} \textbf{\bibinfo{volume}{7}},
  \bibinfo{pages}{401–430} (\bibinfo{year}{2007}).

\bibitem[{\citenamefont{Verstraete et~al.}(2008)\citenamefont{Verstraete, Murg,
  and Cirac}}]{verstraete_matrix_2008}
\bibinfo{author}{\bibfnamefont{F.}~\bibnamefont{Verstraete}},
  \bibinfo{author}{\bibfnamefont{V.}~\bibnamefont{Murg}}, \bibnamefont{and}
  \bibinfo{author}{\bibfnamefont{J.}~\bibnamefont{Cirac}},
  \bibinfo{journal}{Adv. Phys.} \textbf{\bibinfo{volume}{57}},
  \bibinfo{pages}{143} (\bibinfo{year}{2008}).

\bibitem[{\citenamefont{Baumgratz et~al.}(2013)\citenamefont{Baumgratz,
  N\"{u}{\ss}eler, Cramer, and Plenio}}]{baumgratz_scalable_2013}
\bibinfo{author}{\bibfnamefont{T.}~\bibnamefont{Baumgratz}},
  \bibinfo{author}{\bibfnamefont{A.}~\bibnamefont{N\"{u}{\ss}eler}},
  \bibinfo{author}{\bibfnamefont{M.}~\bibnamefont{Cramer}}, \bibnamefont{and}
  \bibinfo{author}{\bibfnamefont{M.~B.} \bibnamefont{Plenio}},
  \bibinfo{journal}{New J Phys.} \textbf{\bibinfo{volume}{15}},
  \bibinfo{pages}{125004} (\bibinfo{year}{2013}).

\bibitem[{\citenamefont{Wright et~al.}(2012)\citenamefont{Wright, Chiao,
  Gevaux, Klopper, Georgescu, and Verberck}}]{Wright_QuantumSimulation_2012}
\bibinfo{editor}{\bibfnamefont{A.}~\bibnamefont{Wright}},
  \bibinfo{editor}{\bibfnamefont{M.}~\bibnamefont{Chiao}},
  \bibinfo{editor}{\bibfnamefont{D.}~\bibnamefont{Gevaux}},
  \bibinfo{editor}{\bibfnamefont{A.}~\bibnamefont{Klopper}},
  \bibinfo{editor}{\bibfnamefont{I.}~\bibnamefont{Georgescu}},
  \bibnamefont{and} \bibinfo{editor}{\bibfnamefont{B.}~\bibnamefont{Verberck}},
  eds., \emph{\bibinfo{title}{Nature Physics Insight -- Quantum Simulation}},
  vol. \bibinfo{volume}{8, no. 4} (\bibinfo{publisher}{Nature Publishing
  Group}, \bibinfo{year}{2012}).

\end{thebibliography}

\end{document}